# C-MORE : the laser guide star wavefront sensor

J.L. Gach*[a,b], D. Boutolleau[b,c], A. Caillat[a], P. Feautrier[b,c], R. Pourcelot[a], E. Stadler[b,c]

[a]Aix Marseille Univ, CNRS, CNES, LAM, Marseille, France;
[b]First light Imaging S.A.S., Europarc Ste Victoire, Route de Valbrillant, 13590 Meyreuil, France;
[c]Univ. Grenoble Alpes, CNRS, IPAG, F-38000 Grenoble, France;

## ABSTRACT

After releasing reference camera solutions in the visible and infrared for natural guide star wavefront sensing with unbeaten performance, we will present the first results of First Light Imaging's C-MORE, the first laser guide star oriented wavefront sensor camera. Within the Opticon WP2 european funded project (INFRAIA 2016-2017, Grant agreement n°730890), which has been set to develop LGS cameras, fast path solutions based on existing sensors had to be explored to provide working-proven cameras to ELT projects ready for the first light schedule. Result of this study, C-MORE is a CMOS based camera with 1600x1100 pixels (9um pitch) and 500 FPS refresh rate. It has been developed to answer most of the needs of future laser based adaptive optics systems (LGS) to be deployed on 20-40m-class telescopes as well as on smaller ones. Using a global shutter architecture, it won't introduce differential temporal errors on the wavefront reconstruction and simplifies the whole command loop. We present the global architecture of the camera, dimensions, weight, interfaces, its main features and measured performance in terms of noise, dark current, quantum efficiency and image quality which are the most important parameters for this application. Because of the very low cost of this solution, this camera can be used also in life-sciences and high end industrial applications, which was also an objective of the Opticon project.

## 1. INTRODUCTION

The idea to use a laser to create an artificial star was introduced by Foy & Labeyrie in 1985 [1] using Rayleigh backscatter light. The concept was then adapted to sodium laser guide stars by Brase at al. [2] later in 1994. This was a perfect solution for 10m-class telescopes, but for the upcoming 40-m class telescopes some new effects appeared due to the pupil size. Indeed, the artificial star takes place at a finite distance (~80 km) and has a quite significant height (up to 10km), therefore when seen from the edges of the pupil, it has an elongated shape. This effect increases when the telescope is pointing away from the zenith. As a first approximation, when looking at the zenith, the angular size of the elongated spot can be expressed by the following formula :

$$\theta = \frac{hr}{H^2} \qquad (1)$$

Where θ is the angular size, h is the sodium layer height, r is the distance to the launch site at the entrance pupil level and H is the sodium layer height, this gives 0.32 arcsecond per meter of distance to the launch laser in the case of 80km sodium layer altitude and 10km thickness. In the case of the 39m E-ELT where the lasers will be launched on the side, this end up to a quite significant 13 arcseconds size, which is the image size obtained on the farthest sub apertures of a shack-hartmann wavefront sensor, like the ones used in the Harmoni instrument [3] (Neichel et al.). Numerous approaches have been studied to mitigate the spot elongation, like the solutions proposed by Ragazzoni et al. [4], Kellner et al. [5], Adkins et al. [6], Gendron [7], Jahn et al. [8] (non-exhaustive). But to date the most straightforward and simple approach is to have a large sensor to sample correctly the spot elongation of the shack hartmann wavefront sensor subaperture. The drawback is mainly a larger cost for the sensor, a larger real time computer, an increase of the necessary transmission bandwidth and more laser optical power because of the spread energy. All these drawbacks are now easy to overcome compared to a more speculative solution. With this approach, the required specs for the sensor are summarized in table 1. ESO carried out a development based on the minimal specs in tems of pixel numbers (Jorden et al. [9]). But it has been proven that spot truncation is a major issue [10] and this development won't cover all the instrument needs. On our side, in the Opticon project we proposed an alternative to this sensor in order to fulfill the instrument requirements as much as possible and to have an alternative as a risk mitigation to the ESO development.

*jeanluc.gach@first-light.fr, www.first-light-imaging.com

Table 1. Specifications of the ELT LGS wavefront sensor cameras. The higher is the weight, the most important the spec is.

| Item | Min spec | Goal spec | weight | Comment |
|---|---|---|---|---|
| Pixel count | 800x800 | 1600x1600 | 3++ | Avoid spot elongation truncation errors |
| Pixel size (μm) | 4 | 34 | 2 | Gives final microlens numerical aperture and physical positioning tolerances |
| Readout speed (Hz) | 300 | 500 | 1 | Computed from simulations of standard ESO wind profile |
| Readout noise (e$^-$) | 4 | 1 | 3 | From simulations of laser return |
| QE (%) | 50% | 90% | 3 | At 586nm. From simulations of laser return |
| Dark current (e$^-$/s/pix) | 200 | 20 | 1 | |
| Defective pixels | 2 by 16x16 clusters | 0 | 1 | |
| Topology | Rolling shutter | Global shutter | 3 | Temporal errors impact. |

## 2. THE C-MORE CAMERA

C-MORE is a global shutter commercial CMOS sensor based camera featuring 1600x1100 pixels of 9 μm and running at 480 to 660 frames per second depending on the digitization depth. Table 2 summarizes the main specs of the camera compared to the instrument needs.

Table 2. Specifications of the ELT LGS wavefront sensor cameras. The higher is the weight, the most important the spec is.

| Item | Min spec | Goal spec | weight | C-more spec |
|---|---|---|---|---|
| Pixel count | 800x800 | 1600x1600 | 3++ | 1600x1100 |
| Pixel size (μm) | 4 | 34 | 2 | 9 |
| Readout speed (Hz) | 300 | 500 | 1 | 480 at 12 bits quantization<br>660 at 8 bit quantization |
| Readout noise (e$^-$) | 4 | 1 | 3 | 2.8 |
| QE (%) | 50% | 90% | 3 | 75% |
| Dark current (e$^-$/s/pix) | 200 | 20 | 1 | 40 |
| Defective pixels | 2 by 16x16 clusters | 0 | 1 | 0 |
| Topology | Rolling shutter | Global shutter | 3 | Global |

The camera will have a CXP-12 interface as well as a 10GigE ethernet which is likely the standard for the future ELT telescopes, a wide input power supply (5-15V) and a gigabit ethernet interface for auxiliary controls if needed. Multiple cameras can be synchronized with 5V tolerant I/Os, which is necessary in the case of laser assisted tomographic adaptive optics where several laser guide star references are used at the same time to probe the atmosphere. Figure 1 shows a CAD view of the camera with its SWaP (Size, Weight and Power) optimized shape of only 53x64x155mm and a weight of a few hundred grams. The sensor will be cooled and thermally stabilized to ensure the performance stability over time. A special care is done in the design to ensure a good thermal management so that all the heat is conducted to the bottom plate of the camera where a heat exchanger (liquid or passive) can be connected. This will ensure a minimal environment disturbance which is a key parameter in the future astronomical instruments.

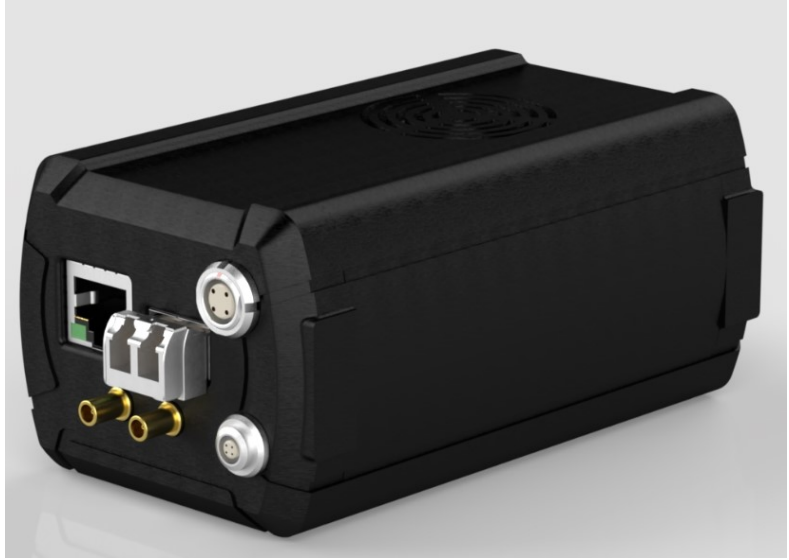

Figure 1: CAD view of the C-MORE camera showing the interfaces at the back.

## 3. PERFORMANCE MEASUREMENT

### 3.1 Test setup

To perform the first electro-optical tests, a test equipment was used. It does not have the full features of the final camera, indeed the sensor is not temperature stabilized, the interfaces are different and the size of this test equipment is larger than the final one (see Figure 2). But the main sensor performance can be measured with this test camera.

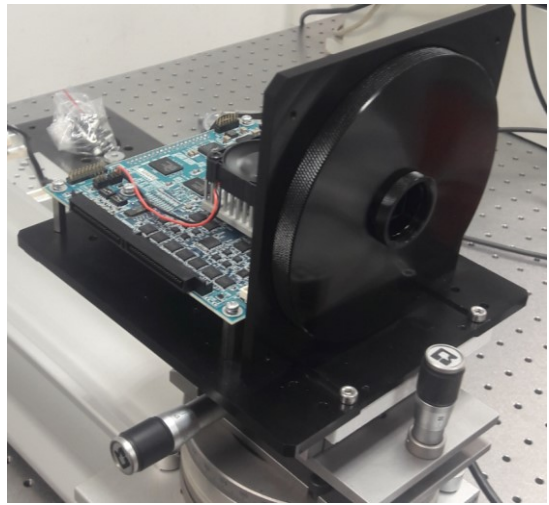

Figure 2: The test camera

All the tests were done using a Labsphere stabilized integrating sphere with a tugsten lamp source.

## 3.2 Photon transfer curve

The photon transfer curve was done in 12 bit conversion mode at 480FPS with the maximum analog gain of 24dB. Figure 3 shows the result of the PTC limited to the non saturating part. The fit shows an excellent linearity and leads to a conversion gain of 0.3 e$^-$/ADU and a temporal readout noise of 2.78 electrons.

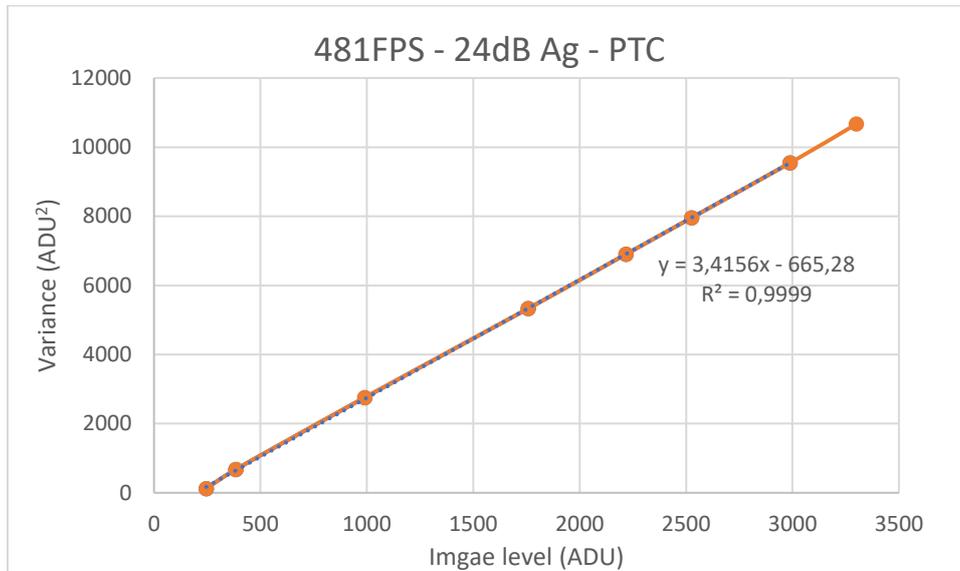

Figure 3: C-MORE Photon transfer curve

## 3.3 Dark current

The dark current was measured by increasing the exposure time in the darkness. Since the sensor temperature was not controlled, it has been measured only for one temperature (+40°C). this gives an upper bound of the dark current. Shows the resulting curve leading to a dark current of 46 e$^-$/s/pix which is already near the goal spec.

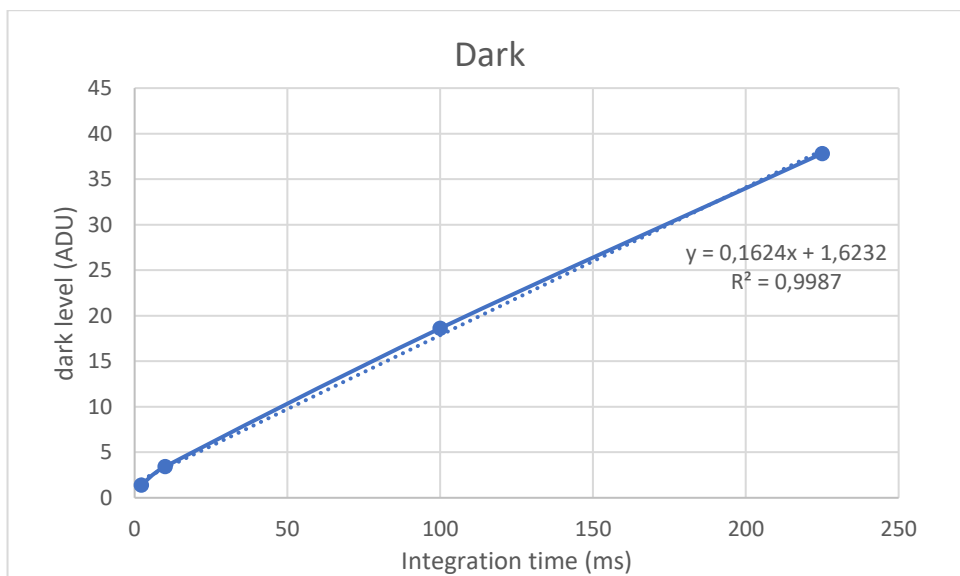

Figure 4: Dark current measurement

## 3.4 Noise performance

The noise histogram is presented in Figure 5 with a mean of 2.78 electrons for the device under test. It shows no long tail even in log scale, which is a common defect of CMOS image sensors. Most of the "noisy" pixels exhibit a RTS beaviour and are evenly distributed on the sensor. Table 3 shows statistics of mean noise over 10 different sensors. This gives a mean noise of 2.77 electrons and a typical dispersion of 0.07 electron from device to device.

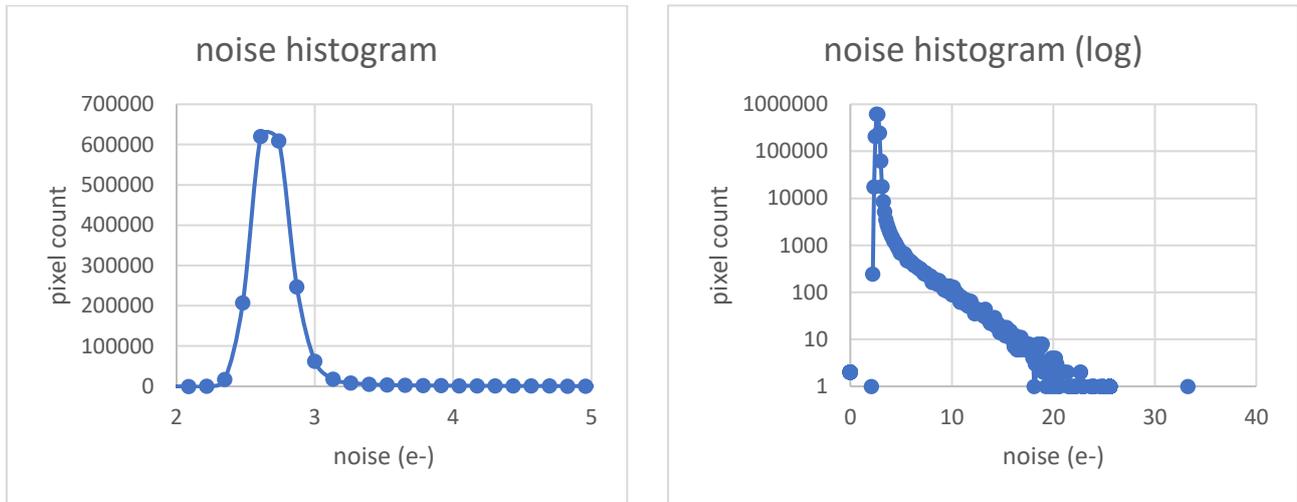

Figure 5: Noise histograms in linear (left) and log (right) scales

Table 3. Readout noise of various devices.

| Device number | Noise (electrons) |
|---|---|
| 1 | 2.84 |
| 2 | 2.69 |
| 3 | 2.90 |
| 4 | 2.65 |
| 5 | 2.67 |
| 6 | 2.90 |
| 7 | 2.74 |
| 8 | 2.73 |
| 9 | 2.80 |
| 10 | 2.81 |

## 3.5 Full well capacity

It is possible to adjust the analog gain to minimize the readout noise. Doing this, the system full well is not limited by the photodiode capacitance but by the ADC conversion scale. By adjusting finely the analog gain, it is then possible to adapt the sensor to the observing conditions, using a lower gain leading to a slightly higher readout noise but a much larger well capacity. Figure 6 summarizes the tradeoff between well capacity and overall readout noise.

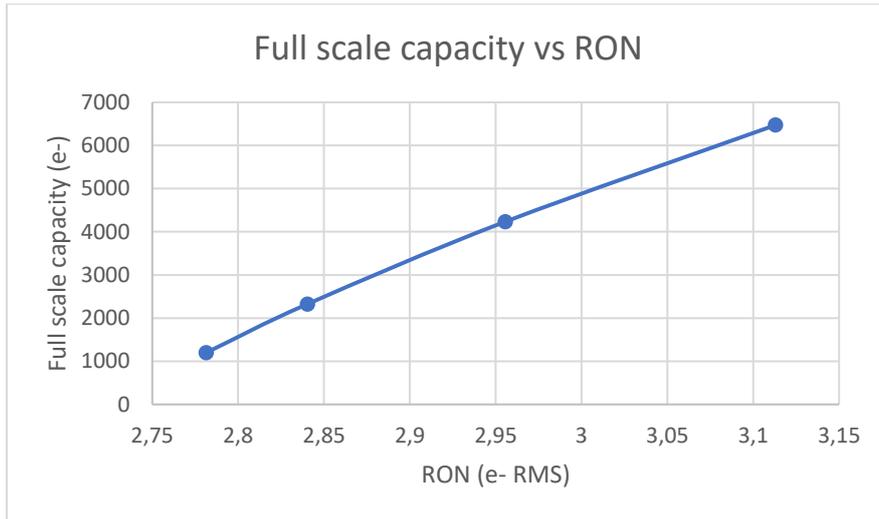

Figure 6: Well capacity vs readout noise achievable by adjusting the analog gain

### 3.6 Quantum efficiency

The quantum efficiency given from the sensor manufacturer's data is shown in Figure 7. Fortunately, it peaks at ~75% at the 586nm sodium wavelength.

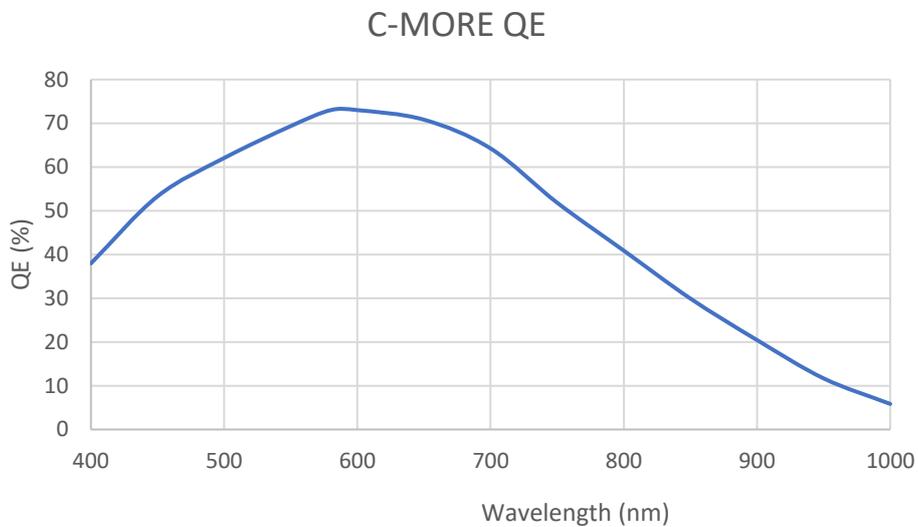

Figure 7: C-MORE QE vs wavelength from sensor manufacturer's data with a maxiùum of 75% at 586 nm

The used sensor is a front side illuminated sensor with microlens array to enhance the overall QE. However this has the drawback of having a maximal acceptance angle where the effective quantum efficiency decreases. Because of the pixel geometry, this angle is not necessarily symmetric nor the same in X and Y directions across the pixel. To make the acceptance measurements we've used the MITHIC bench (see N'Diaye et al. 2012[11] & Vigan et al. 2016[12]) modified with a supercontinuum source and a visible transmission filter from 400 to 600nm to avoid light interferences present with the original 650nm laser source. This bench provides a highly collimated beam of 5mm with wavefront correction. Figure 8 shows the test setup, with the test camera mounted on a dual axis rotation stage and elevation system to intercept the MITHIC beam.

The results of the acceptance tests are presented in and compared to the sensor manufacturer's data. The measurements show an excellent accordance between them. Considering the Y axis as the most impacted by the acceptance angle, a beam slower than F/3 will affect marginally the sensor's overall QE which is compatible with an adaptive optics system on a 40m telescope.

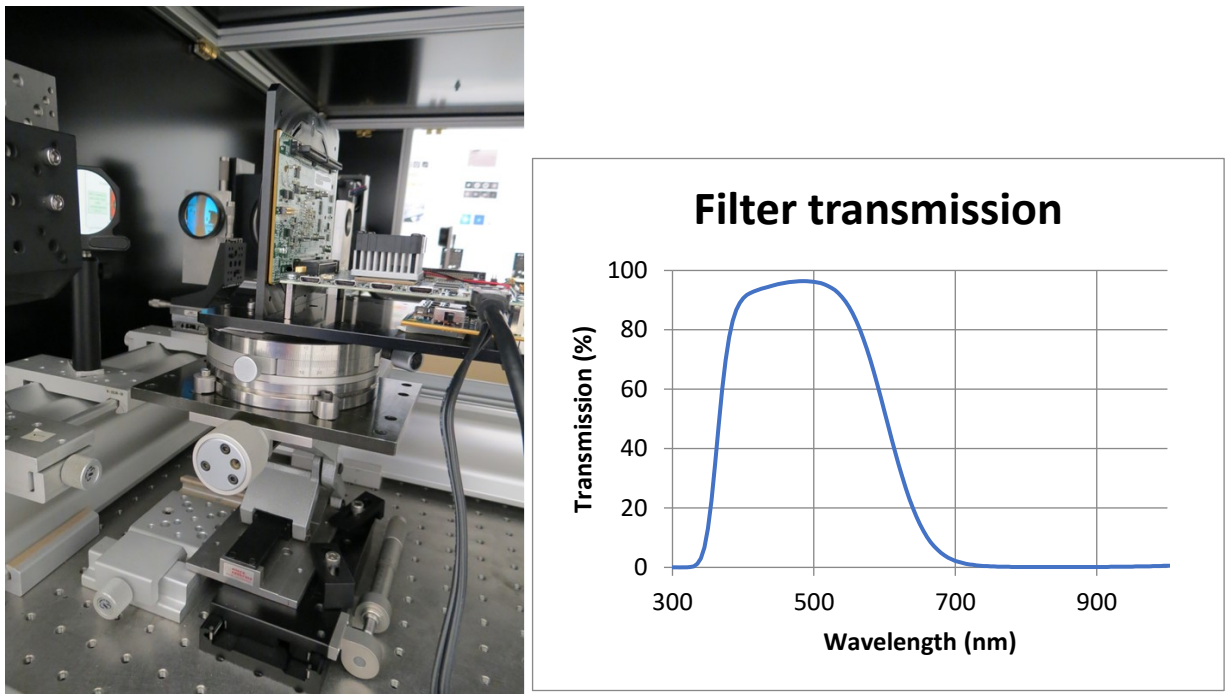

Figure 8: The test camera installed on the MITHIC bench (left) and the filter transmission (right).

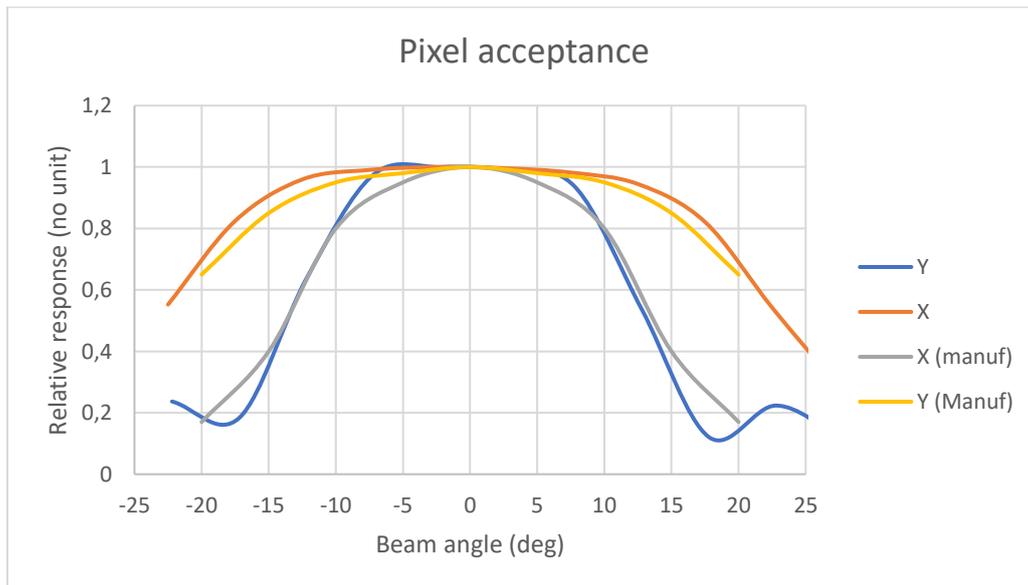

Figure 9: measured pixel acceptance in X a Y and comparison with the sensor manufacturer's data

### 3.7 Readout modes

Taking the benefit of the numerous readout possibilities of the CMOS sensors it is also possible to use region of interest (ROI) to increase the frame rate of the sensor. The frame rate scales with the number of lines dead out, therefore a 256x256 window would lead to 2067 FPS readout which is exactly the same frame rate as the OCAM2 camera [13][14] which is currently used as a NGS or LGS wavefront sensor on 10m class telescopes, where the spot elongation is not an issue. A 128x128 ROI would lead to more than 4000FPS which is ideal for small space awareness and surveillance telescopes of the 1 to 4m class (d'Orgeville et al. 2014[15], Grosse et al. 2017[16] or Bennet at al. 2014[17]) where the apparent wind speed is very high and the number of subapertures is small due to the small entrance pupil size.

## 4. CONCLUSION

We showed a very promising path to solve the spot elongation truncation issue on 40m class shack-hartmann based laser guide star adaptive optics systems, using commercially available components. A specialized camera is under development at First Light Imaging to fulfill most of the needs of these systems and will be available soon. As it uses already developed sensors the cost of such devices will be much lower than the other developments carried out. Because of this and the various readout modes it is also a very good candidate for small or even very small telescopes LGS systems. Finally because of its outstanding performance especially in terms of noise these cameras could also replace in the future the very expensive EMCCD based cameras even of natural guide stars.